\documentclass[letterpaper]{article} % DO NOT CHANGE THIS
\usepackage{aaai25}  % DO NOT CHANGE THIS
\usepackage{times}  % DO NOT CHANGE THIS
\usepackage{helvet}  % DO NOT CHANGE THIS
\usepackage{courier}  % DO NOT CHANGE THIS
\usepackage[hyphens]{url}  % DO NOT CHANGE THIS
\usepackage{graphicx} % DO NOT CHANGE THIS
\def\UrlFont{\rm}  % DO NOT CHANGE THIS
\usepackage{natbib}  % DO NOT CHANGE THIS AND DO NOT ADD ANY OPTIONS TO IT
\usepackage{caption} % DO NOT CHANGE THIS AND DO NOT ADD ANY OPTIONS TO IT
\usepackage{algorithm}
\usepackage{algorithmic}
\usepackage{amssymb}
\usepackage{amsmath}
\usepackage{booktabs}
\usepackage{comment}

\DeclareMathOperator*{\argmin}{arg\,min}

\usepackage{newfloat}
\usepackage{listings}
\DeclareCaptionStyle{ruled}{labelfont=normalfont,labelsep=colon,strut=off} % DO NOT CHANGE THIS
\floatstyle{ruled}
\newfloat{listing}{tb}{lst}{}
\floatname{listing}{Listing}
\title{An Uncertainty-Aware Data-Driven Predictive Controller for Hybrid Power Plants}
\author{
   Manavendra Desai, Himanshu Sharma, Sayak Mukherjee, Sonja Glavaski
}
\affiliations{
    Pacific Northwest National Laboratory
    Richland, WA, 99354,USA\\
    \{manavendrabalwant.desai, himanshu.sharma,sayak.mukherjee, sonja.glavaski\}@pnnl.gov
}

\usepackage{bibentry}
\begin{document}

\maketitle

\begin{abstract}
 Given the advancements in data-driven modeling for complex engineering and scientific applications, this work utilizes a data-driven predictive control method, namely \textit{subspace predictive control}, to coordinate hybrid power plant components and meet a desired power demand despite the presence of weather uncertainties. An uncertainty-aware data-driven predictive controller is proposed and its potential is analyzed using real-world electricity demand profiles. For the analysis, a hybrid power plant with a wind, solar, and co-located energy storage capacity of 4 MW each, is considered. The analysis shows that the predictive controller can track a real-world inspired electricity demand profile despite the presence of weather-induced uncertainties and be an intelligent forecaster for HPP performance.
\end{abstract}

\section{Introduction}
% Relevant References for Hybrid pkatn :
% https://emp.lbl.gov/hybrid
\noindent \textbf{Background:} The world is rapidly transitioning towards energy diversification to meet energy demands and reduce energy cost. Advances in renewable energy generation technologies, such as solar panels and wind turbines, along with energy storage solutions, have heightened the focus on their reliable and sustainable integration with conventional energy systems \cite{bolinger2023hybrid}. Hybrid power plants (HPP) have emerged as a promising solution, combining two or more technologies, which may include wind turbines, photovoltaic (PV) or concentrated solar power systems, energy storage, geothermal power, hydropower, biomass, natural gas, oil, coal, or nuclear power. In this study, we focus on HPPs designed for electricity production, specifically those leveraging renewable generation, such as wind and solar, combined with energy storage \cite{bolinger2023hybrid}. In the U.S., as of the end of 2023, solar PV and energy storage plants have led the market, with a growing presence of wind, solar, and energy storage combinations. Additionally, there is strong interest in hybridization among developers of renewable energy systems, evidenced by an increase in interconnection requests for HPPs.

Developers are also exploring advancements in power electronics for integrating renewable assets like solar, wind, and energy storage, to mimic capabilities of traditional power plants in terms of their capacity value, dispatchability, ancillary services, and reliability. The heightened penetration of renewable energy-dominant systems also motivates researchers to reconsider revenue streams for wind-based HPPs, comparable to conventional power plants \cite{dykes2020opportunities}. Additionally, this has inspired research in simultaneously optimizing for system design and operation (i.e control co-design) \cite{garcia2019control,herber2017advances, bird2024set, SharmaGraph} to ensure long term efficiency and reliable operation of the system. Furthermore, new regulations for grid operators and agencies are being assessed to mandate roles of renewable-energy generators in fortifying the power grid during contingencies \cite{li2023review}. Consequently, the design objectives for HPPs are shifting from merely reducing the levelized cost of energy to identifying income streams that maximize revenue through participation in time-varying energy pricing markets, ancillary services, and capacity markets. Further discussions on HPPs are available in \cite{dykes2020opportunities}.

\noindent \textbf{Motivation:} The ability of HPPs to function comparably to traditional power-generation plants is often compromised by the uncertainty and variability inherent in wind, solar, and energy storage technologies. This presents both a challenge and an opportunity for developers of HPPs to innovate control systems that deliver economic benefits akin to traditional power plants. Addressing this uncertainty can enhance the reliability of and decision-making in HPPs. 
% This study proposes to examine the role of accounting for uncertainty in the design of data-driven predictive controllers for hybrid power plants. 
With the advancement and proliferation of sensing technology in complex physical systems, data-driven predictive controller designs are anticipated to offer the necessary flexibility for HPP deployment, surpassing that of model-based controller designs. For hybrid systems encompassing multiple assets, such as wind, solar, and storage, capturing the dynamic operations of the plant is crucial for effective controller design.

\noindent \textbf{Related work:} In the age of "big data", data-driven modeling and decision-making for various engineering and scientific applications intersect significantly with the field of control theory \cite{annaswamy2024control}. Both fields mutually enhance their applicability to complex engineering challenges \cite{brunton2022data}. Recent advances in model-free decision-making have spurred significant developments in reinforcement learning and adaptive dynamic programming, both used in Markov decision processes (MDPs) \cite{sutton2018reinforcement, powell2007approximate}, as well as in control-theoretic state-space approaches with rigorous guarantees \cite{vrabie2009adaptive, jiang2012computational, mukherjee2021reduced}. Among these data-driven control methodologies, behavioral system theory has shown considerable promise, as highlighted in works such as \cite{de2019formulas, van2020data}. This progress has led the development of predictive extensions, including methods like data-enabled predictive control \cite{coulson2019data} and advancements in subspace predictive control \cite{fiedler2021relationship}. A comprehensive overview of data-driven control from a control theoretical perspective is provided by \cite{prag2022toward}, and for process control applications, by \cite{tang2022data}. A systematic exploration of how data-driven models and approaches can leverage cutting-edge machine learning methods to identify system dynamics crucial for control design is provided in \cite{brunton2022data}. Lastly, the integration of learned components into traditional control techniques, or deriving control actions directly from data, is increasingly popular due to advances in computing power and data acquisition. Reinforcement learning-based approaches have been identified as promising control solutions across various engineering applications, with detailed reviews presented by \cite{wang2022deep,hewing2020learning}.
\noindent \textbf{Objective and main contributions:} This paper proposes, and investigates the potential of, an uncertainty-aware data-driven predictive controller as an intelligent decision-maker and supervisor for a hybrid power plant.
In this work, intelligence of the controller is considered and demonstrated in the context of its potential ability to a) encode data-driven HPP dynamics, b) provide forecasts of the power output of each component of the HPP, and c) anticipate and control the state of charge/discharge of the energy storage device in the event of peak demands from the electricity power grid.  

Next, a simulation study that utilizes real-world electricity demand profiles is presented and the performance of the controller is analyzed. Lastly, open research inquiries to make the decisions of the controller trustworthy and explainable, are discussed. 

The rest of this paper is organized as follows. Section \ref{sec:method} introduces the HPP considered in this work and the design and application of the uncertainty-aware data-driven predictive controller. Results of the simulation study and related discussions are provided in Sections \ref{sec:results} and \ref{sec:disc}, respectively. Finally, concluding remarks are made in Section \ref{sec:conc}.
\begin{figure}
    \centering
\includegraphics[width=\linewidth]{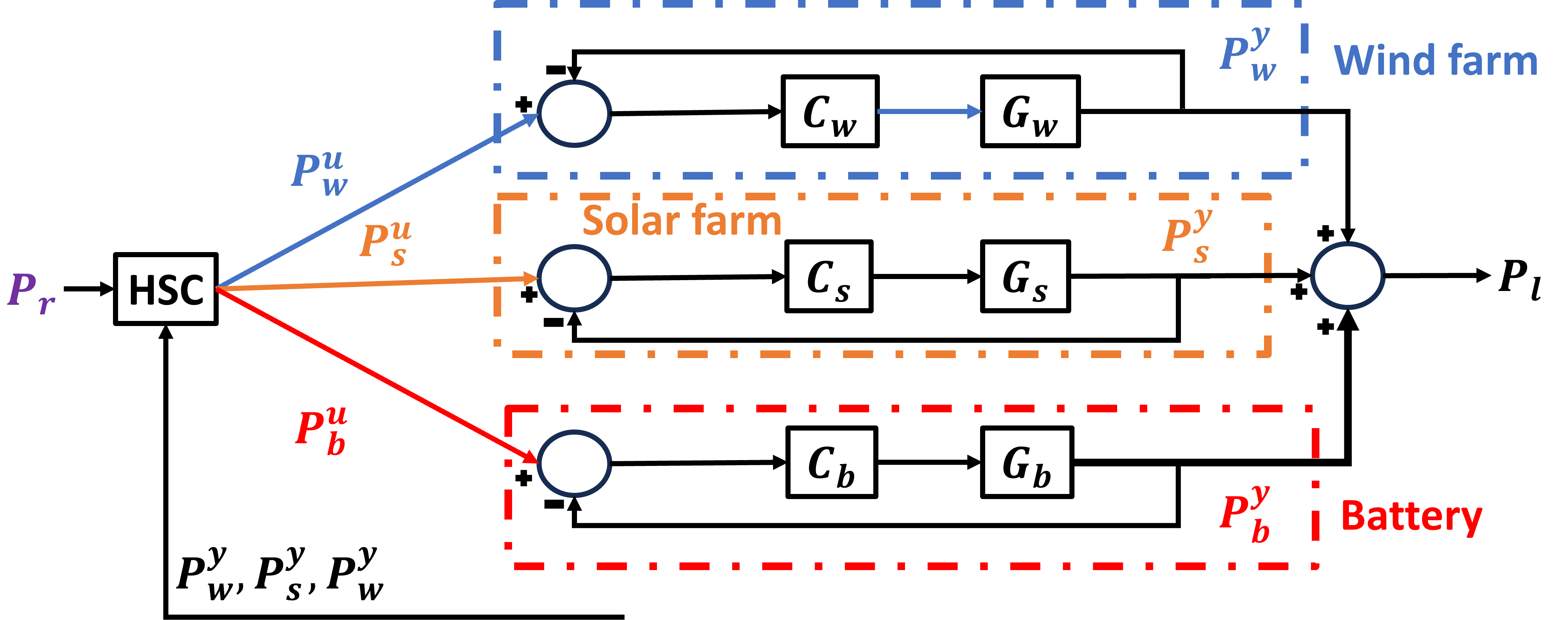}
    \caption{The control scheme of the hybrid power plant (HPP) considered in this work. The HPP supervisory controller ($HSC$) attempts to track a load reference $P_r$ by selecting optimal setpoints (superscript $u$) for the underlying plant components. Measured outputs (superscript $y$) are summed to obtain the overall HPP output $P_l$. $\mathbf{C}$ and $\mathbf{G}$ respectively represent controller and system dynamics, for each plant component.}
    \label{fig:hsc_control_diag}
\end{figure}

\section{Methodology}
\label{sec:method}
% We present the details about the simulation environment, dataset and the predictive controller design for the hybrid energy power plant.
\noindent \textbf{Hybrid power plant:} This work considers an HPP that comprises a wind farm, solar farm, and battery as an energy storage device. Figure \ref{fig:hsc_control_diag} shows the \textit{supervisory} control scheme for HPP control used in this work. Setpoints are denoted by $u$, and measured outputs are denoted by $y$. The HPP supervisory controller ($HSC$) selects optimal set-points for tracking an electricity load power reference $P_r$. The subscripts $w$, $s$, and $b$, denote the relation to the wind farm, solar farm, and battery, respectively. The wind and solar farms generate power outputs, $P^{y}_w$ and $P^{y}_s$, respectively. Excess power produced (i.e., when $P^{y}_w + P^{y}_s > P_r$), is used to charge the battery ($P^{u}_b<0$). In the event of a deficit (i.e., when $P^{y}_w + P^{y}_s < P_r$), the battery is discharged (i.e., $P^{u}_b>0$) to support load-tracking. Each plant energy generation component (solar PV, turbines) comprises a dedicated and well-tuned proportional-integral-derivative (PID) controller, $\mathbf{C}$, for tracking the respective setpoints - $P^{u}_w$, $P^{u}_s$ and $P^{u}_b$. The wind farm, solar farm, and battery, for tracking the reference load, are approximately modeled as first-order systems, $\mathbf{G}$. Note that $P^y_w$ and $P^y_s$ are limited by the wind and solar energy available, thus making weather forecasts and the associated uncertainties in available wind and solar energy critical to HPP performance.

\noindent \textbf{Subspace predictive control:} 
The data-driven predictive controller introduced in this work is based on the method of \textit{subspace predictive control} \cite{favoreel1999spc}. Subspace predictive control (SPC) is a parameter-free technique that identifies a multi-step prediction model of a system using only previously measured input-output data and uses this model to formulate an optimal control problem.

Consider the measured input and output data, $U$ and $Y$, respectively, to comprise $T$ trajectories of a multi-input and multi-output dynamic system. Each trajectory is $L=T_{ini}+N$ samples long. $N$ is the length of the prediction horizon of the controller. $T_{ini}$ is the number of the most recent measurements of $U$ and $Y$ of relevance. Let $u$ $\epsilon$ $\mathbb{R}^{m}$ and $y$ $\epsilon$ $\mathbb{R}^{p}$. Therefore, $U$ and $Y$ can be expressed as $U=[u^1_L~u^2_L~...~u^{T}_L]$ and $Y=[y^1_L~y^2_L~...~y^{T}_L]$, where $u^j_L=[(u^j_1)^{\intercal}~(u^j_2)^{\intercal}~...~(u^j_L)^{\intercal}]^{\intercal}$  $\epsilon$ $\mathbb{R}^{Lm}$ and $y^j_L=[(y^j_1)^{\intercal}~(y^j_2)^{\intercal}~...~(y^j_L)^{\intercal}]^{\intercal}$ $\epsilon$ $\mathbb{R}^{Lp}$. Fulfilling Assumptions 1, 2, and 3, discussed in \cite{fiedler2021relationship}, ensures that the values of $T_{ini}$, $N$ and $T$, satisfy observability requirements for data-driven predictive control, and make $U$ persistently exciting. Splitting each of $U$ and $Y$ as $U=[~U_{T_{ini}}^{\intercal}~~~U_N^{\intercal}~]^{\intercal}$ and $U=[~Y_{T_{ini}}^{\intercal}~~~Y_N^{\intercal}~]^{\intercal}$, a matrix $S^{*}$ can be calculated as

\begin{equation}
    S^{*} = \argmin_{S} ||S \underbrace{\begin{bmatrix}
       Y_{T_{ini}} \\ U_{T_{ini}} \\ U_N
    \end{bmatrix}}_M - Y_N||^2_F, 
    \label{eq:Sstar}
\end{equation}
to give $S^{*}=Y_NM^{\dagger}$, where $\dagger$ denotes a Moore-Penrose pseudo-inverse. Here, $U_{T_{ini}}$ $\epsilon$ $\mathbb{R}^{T_{ini}m \times T}$ and $Y_{T_{ini}}$ $\epsilon$ $\mathbb{R}^{T_{ini}p \times T}$. Additionally, $U_N$ $\epsilon$ $\mathbb{R}^{Nm \times T}$ and $Y_{N}$ $\epsilon$ $\mathbb{R}^{Np \times T}$.

If $U$ and $Y$ are sufficiently rich to cover diverse trajectories or \textit{behaviors} of the HPP system, the matrix $S^{*}$ can be used to represent any single input-output trajectory $y_{T_{ini}}$, $y_N$, $u_{T_{ini}}$, $u_N$, as
\begin{equation}
    y_N=S^{*} \begin{bmatrix}
       y_{T_{ini}} \\ u_{T_{ini}} \\ u_N
    \end{bmatrix},
    \label{eq:y_N}
\end{equation}
where $y_{T_{ini}}$ $\epsilon$ $\mathbb{R}^{T_{ini}p}$, $u_{T_{ini}}$ $\epsilon$ $\mathbb{R}^{T_{ini}m}$, $u_{N}$ $\epsilon$ $\mathbb{R}^{mN}$, and $y_{N}$ $\epsilon$ $\mathbb{R}^{pN}$. Therefore, given the past $T_{ini}$ sequence of input-output measurement data, \eqref{eq:Sstar} provides a relation between inputs and output over a future horizon of $N$ samples. Since $S^{*}$ is calculated using diverse trajectories of the HPP system, the trajectory of ($u_N$, $y_N$) is expected to be admissible by the HPP dynamics.
\begin{figure}[tb]
    \centering
\includegraphics[width=\linewidth]{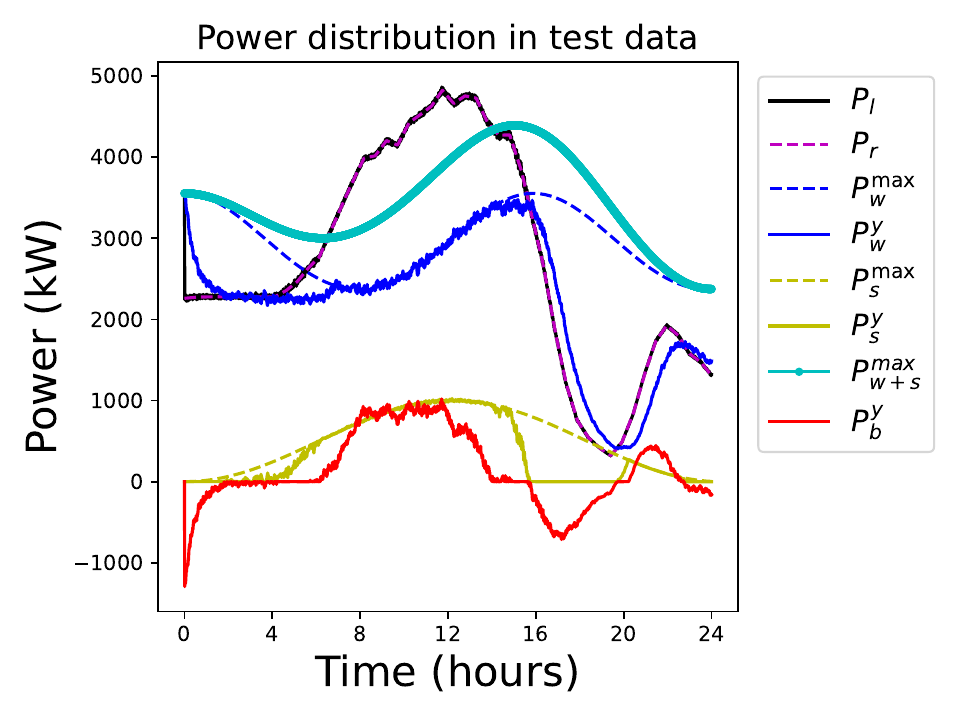}
    \caption{A sample load-tracking result. A constrained feedback-optimization based controller generates input-output data for multiple load references $P_r$ \cite{ortmann2022online}.}
    \label{fig:Test_load}
\end{figure}

\noindent \textbf{Data curation:} In the simulation study presented in this work, the HPP is considered to serve a region with electricity demands equivalent to the combined day-long power profiles of the Hornsea 1 wind farm, and the solar farms of PES Region 10 in East England \cite{heftdata}. 

The input-output data required to determine $S^{*}$ in \eqref{eq:Sstar} is obtained by tracking day-long demand profiles of multiple days, spread over a year. For data collection, a constrained feedback-optimization (FO) based \textit{reactive} controller is utilized as the $HSC$ of the HPP in Figure \ref{fig:hsc_control_diag}. The FO uses a single output measurement (i.e., $y$) of the most recently sampled interval to optimally select setpoints (i.e., $u$) for the next control action \cite{ortmann2022online}. Perfect setpoint-tracking by the inner control-loops of the HPP, and completely certain weather forecasts, are assumed. A noise-to-signal ratio of 0.02 is considered while recording $y$ and $u$.  Figure \ref{fig:Test_load} shows a sample result for tracking an electricity load-reference $P_r$ using the FO controller. Here, the power outputs of the wind farm (i.e., $P^y_w$) and solar farm (i.e., $P^y_s$) are upper bounded by the completely certain and maximum available wind (i.e., $P^{\max}_w$) and solar energy (i.e., $P^{\max}_s$).

\noindent \textbf{Uncertainty-aware predictive controller:} This work introduces an SPC-based data-driven predictive controller for supervisory control of an HPP (see figure \ref{fig:hsc_control_diag}), and analyzes its decision-making in the presence of uncertain weather conditions, particularly wind speed and the wind energy available to the wind farm. The controller is framed as follows.
\begin{eqnarray}
\nonumber && \argmin_{u_N, y_N, \sigma_u, \sigma_y} \sum_{k=1}^{N} \left\{\frac{1}{2}Q_r{J_T^k}^2 + \lambda {\Delta P^{y,k}_w}^2 + \lambda {\Delta P^{u,k}_w}^2 \right\} \\ && + 
 \nonumber \lambda_u||\bf{\sigma_u}||^2 + \lambda_y||\bf{\sigma_y}||^2 \\
&&\text{s.t.} ~ J^k_T = P^{k}_r-P^{y,k}_w-P^{y,k}_s \label{eq:J}\\ &&
~~~~~~y_N = S^*\begin{bmatrix}
    y_{T_{ini}} \\ u_{T_{ini}} \\ u_N \\
\end{bmatrix} + \begin{bmatrix}
    \bf{\sigma_y} \\ \bf{\sigma_u} \\ \bf{0}
\end{bmatrix} \label{eq:FL} \\&&
~~~P^{u,k}_b=P^{k}_r-P^{y,k}_w-P^{y,k}_s \label{eq:Pbd}\\&&
~~~0 \le P^{u,k}_w, P^{y,k}_w \le q_{P^{\max}_{w}} \label{ineq:Pwmxmn}\\&&
~~~0 \le P^{u,k}_s, P^{y,k}_s \le P^{\max}_{s} \label{ineq:Psmxmn}\\&& \label{ineq:Pbmxmn}
~~~P^{\min}_{b} \le P^{u,k}_b, P^{y,k}_b \le P^{\max}_{b}.
\end{eqnarray}
Here, $u_{N}$ is a solution trajectory of setpoints sent by the $HSC$ to the HPP components. $y_{N}$ is the prediction of HPP outputs over the horizon of $N$ samples. $Q_r$ is a penalty on the sample-wise tracking cost $J^{k}_T$ in \eqref{eq:J}. $\lambda$ penalizes large changes in setpoint and power output of the wind farm. \eqref{eq:FL} constrains $u_{N}$ and $y_{N}$ to within the admissible system trajectories, as shown in \eqref{eq:y_N}. $\bf{\sigma_u}$ $\epsilon$ $\mathbb{R}^{T_{ini}m}$ and $\bf{\sigma_y}$ $\epsilon$ $\mathbb{R}^{T_{ini}p}$ are heavily penalized relaxations added in \eqref{eq:FL} to add robustness against measurement noise \cite{coulson2019data}. In \eqref{eq:Pbd}, the setpoint for the battery is constrained to autonomously charge the excess ($P^{u,k}_b<0$), or discharge to support a deficit ($P^{u,k}_b>0$), in the combined output of the wind and solar farms. Constraints \eqref{ineq:Pwmxmn}-\eqref{ineq:Pbmxmn} set bounds on the admissible setpoints to, and power outputs of, the HPP components. $q_{P^{\max}_w}$ is an upper bound for the available wind power in the presence of uncertain forecasts for wind speed. The uncertainty is assumed to have a normal distribution and $q_{{P^{\max}_w}}$ is set using a suitable quantile value. Therefore, \eqref{ineq:Pwmxmn} is an individual probabilistic constraint and adds uncertainty-awareness to the predictive controller. In this work, $N=20$, $T_{ini}=20$, $L=40$, and $T=1000$. Additionally, $p=3$ and $m=3$.

\begin{figure}[tb]
    \centering
\includegraphics[width=\linewidth]{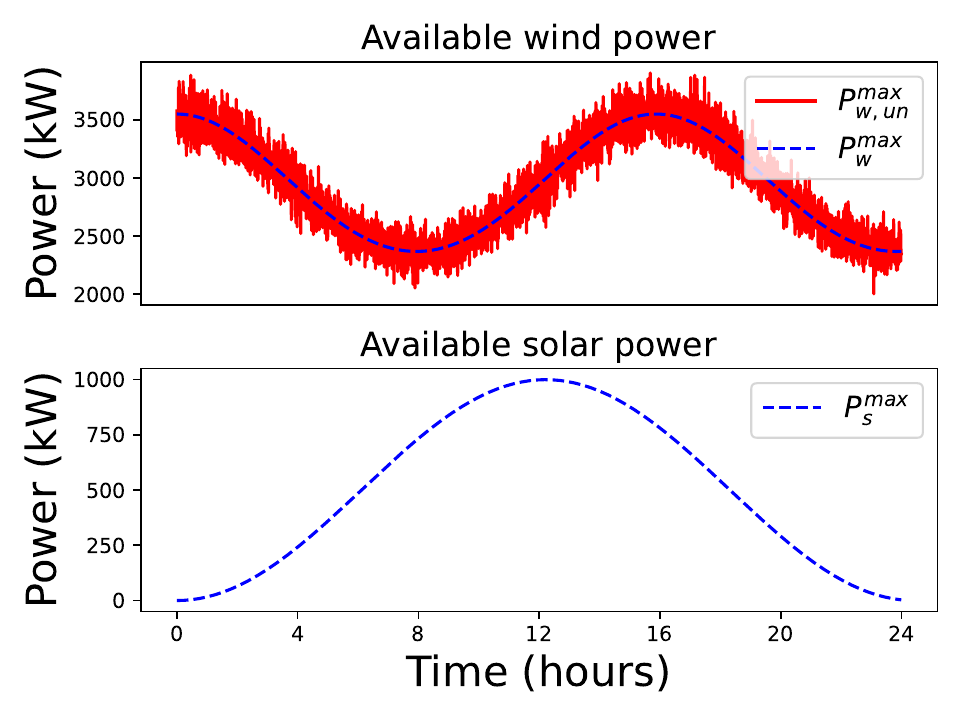}
    \caption{Day-long forecasts for available wind and solar power to be used in the simulation study in Section \ref{sec:results}. Uncertainty added to the wind-speed translates to uncertainty in available wind power ($P^{\max}_{w,un}$).}
    \label{fig:power_curves}
\end{figure}

\noindent \textbf{Simulation setup:} The HPP in Figure \ref{fig:hsc_control_diag} is simulated using the uncertainty-aware data-driven predictive controller as the $HSC$. The wind and solar farms, and the battery, have a rated capacity of 4 MW each. The battery is assumed to have a sufficient state-of-charge at all times and its cycling limits are not considered. The electricity demand profiles in \cite{heftdata} are linearly scaled to the order of five megawatts to be consistent with the rated output of the HPP. The available wind power (i.e., $P^{\max}_w$) and solar power (i.e., $P^{\max}_s$) are functions of the wind speed and solar irradiance, respectively (see Figure \ref{fig:power_curves}). For this study, their profiles are assumed sinusoidal and are generated synthetically. To test the uncertainty-awareness of the predictive controller, a normally distributed uncertainty in wind speed, of mean 0 m/s and standard deviation 0.1 m/s, is applied. Therefore, $q_{P^{\max}_w}=\mu_{P^{\max}_w}+q_w\sigma_{P^{\max}_w}$ in \eqref{ineq:Pwmxmn}. Here, the quantile $q_w=-0.4$, and $\mu_{P^{\max}_w}$ and $\sigma_{P^{\max}_w}$ are mean and standard deviation of the uncertain wind power profile. Figure \ref{fig:Test_load} shows the load-profile $P_r$ used as the reference to the predictive controller. 

The optimization problem in \eqref{eq:J}-\eqref{ineq:Pbmxmn} is solved using the NLP solver from CasADi on a laptop with an 11th Gen Intel(R) Core(TM) i7-1185G7 3.00GHz processor and 16GB RAM. The mean solve time was 0.3 s, much lower than the sampling interval of 20 s of the predictive controller. Lastly, given this sampling interval, the predictive controller solves for the next 20$N$ seconds using measurement data from the past 20 $T_{ini}$ seconds.
\begin{figure}[tb]
    \centering
\includegraphics[width=\linewidth]{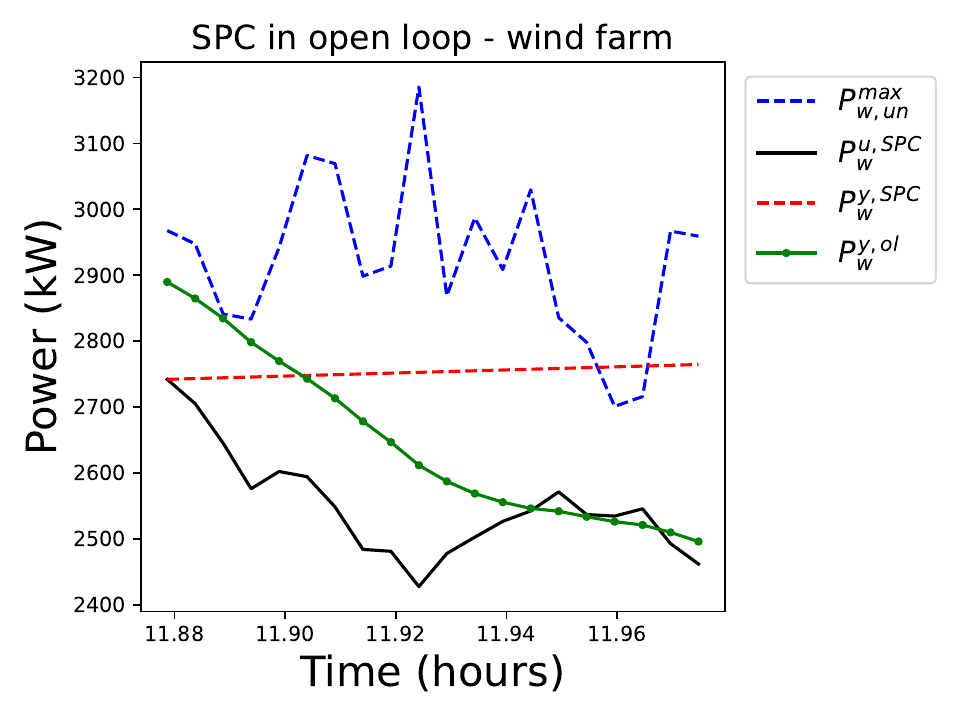}
    \caption{Open-loop uncertainty-aware setpoint selection and tracking for the wind farm. $P^{y,ol}_w$ is the \textit{open-loop} response of the wind farm to setpoints $P^{u,SPC}_w$ of the predictive controller, in the presence of uncertainty in available wind power $P^{max}_{w,un}$. The error of the predicted output $P^{y,SPC}_w$  with respect to $P^{y,ol}_w$, normalized by the mean of $P^{y,SPC}_w$, is 6.5\%.}
    \label{fig:SPC_OL_un_wind}
\end{figure}
\begin{figure}[tb]
    \centering
\includegraphics[width=\linewidth]{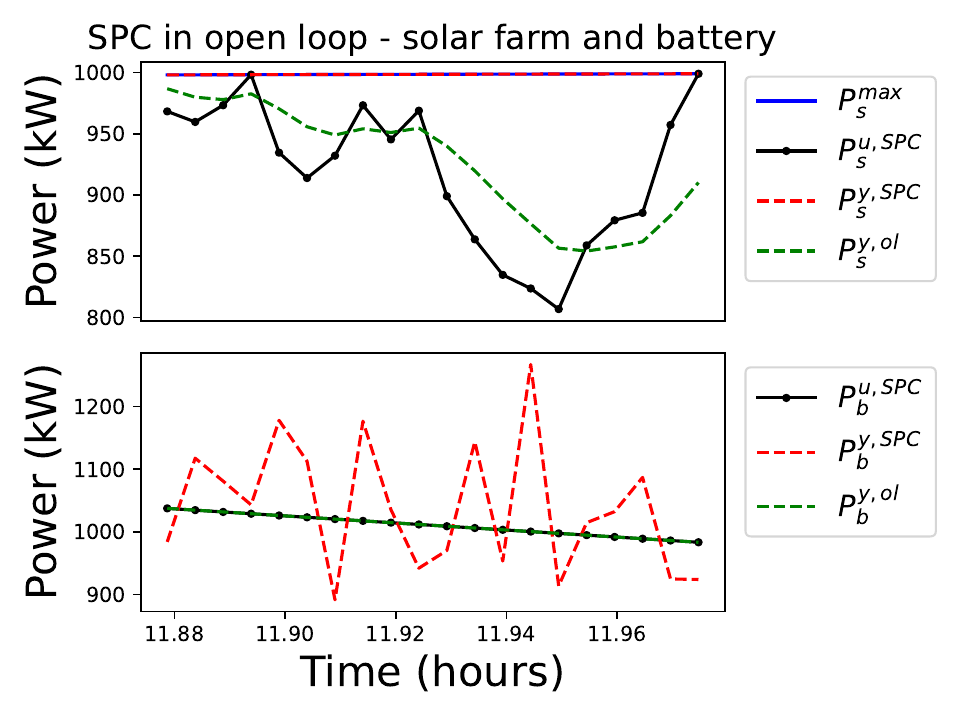}
    \caption{Open-loop setpoint selection and tracking for the solar farm and battery. $P^{y,ol}_s$ and $P^{y,ol}_b$ are respective \textit{open-loop} responses of the solar farm  and battery to setpoints $P^{u,SPC}_s$ and $P^{u,SPC}_b$ of the predictive controller. The error of the predicted solar farm output $P^{y,SPC}_s$  with respect to $P^{y,ol}_s$, normalized by the mean of $P^{y,SPC}_s$, is 8.5\%. For the battery, this value is 10\%.}
    \label{fig:SPC_OL_un_Solar_Batt}
\end{figure}
\begin{figure}[tb]
    \centering
\includegraphics[width=\linewidth]{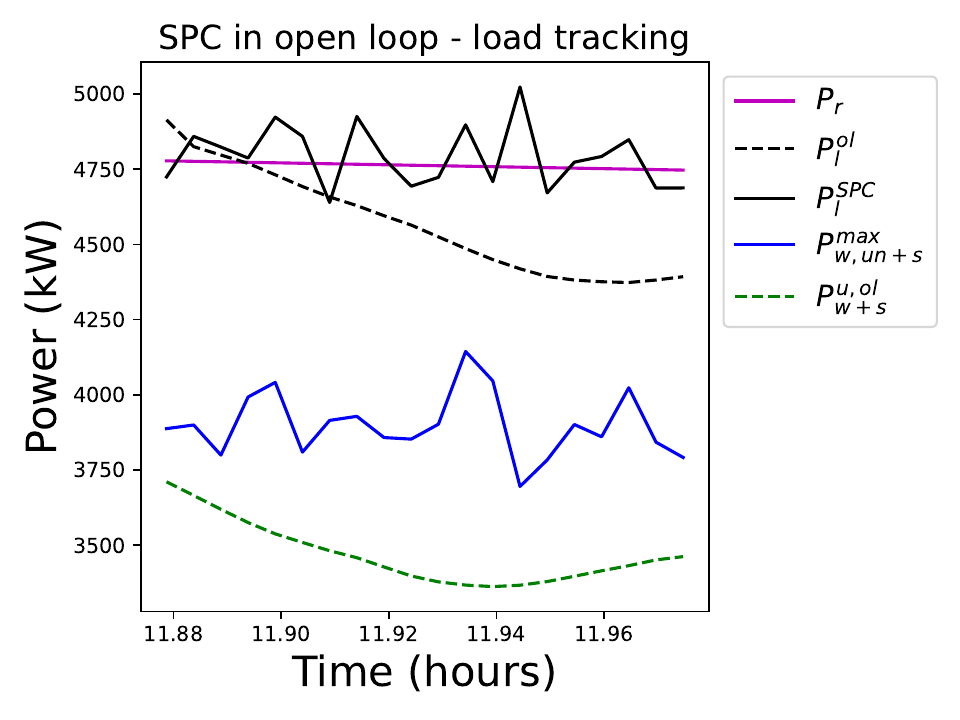}
    \caption{Open-loop uncertainty-aware load tracking for the hybrid power plant in Figure \ref{fig:hsc_control_diag}. $P^{ol}_l$ is the \textit{open-loop} output of the hybrid power plant in response to setpoints of the predictive controller, in the presence of uncertainty in available wind power. The error of the predicted HPP output, $P^{SPC}_l=P^{y,SPC}_w+P^{y,SPC}_s+P^{y,SPC}_b$, with respect to $P^{ol}_l$, and normalized by the mean of $P_r$ over the prediction horizon, is 6.1\%. The combined setpoints to the wind and solar farm, $P^{u,ol}_{w+s}$, obey the probabilistic constraint in \eqref{ineq:Pwmxmn} to remain below the available wind and solar energy, $P^{\max}_{w,un+s}$. }
    \label{fig:SPC_OL_un_load}
\end{figure}
\section{Results}
\label{sec:results}

\noindent \textbf{Open-loop control:} The potential of the uncertainty-aware data-driven predictive controller is tested by first applying its chosen setpoints, $P^{u,SPC}_w$, $P^{u,SPC}_s$, and $P^{u,SPC}_b$, to the HPP of Figure \ref{fig:hsc_control_diag}, in an open-loop fashion, over the prediction horizon of $N$ samples. Therefore, in open-loop control, the entire sequence of setpoints $u_N$, solved by the predictive controller in \eqref{eq:J}-\eqref{ineq:Pbmxmn}, is applied to the HPP without any feedback of the measured HPP output to the controller for a revised calculation of $u_N$. The HPP output obtained is the open-loop response. The simulation is conducted over $N=20$ samples and with the initial sample at 11.88 hours. At this time of the day, Figure \ref{fig:Test_load} shows maximum utilization of wind and solar energy, albeit in the absence of uncertainty. Therefore, such an initial condition can be used to test the uncertainty-awareness of the predictive controller since $P^{u,k}_w$ and $P^{y,k}_w$ are expected to approach an uncertain $P^{\max}_w$. 

Figure \ref{fig:SPC_OL_un_wind} plots the open-loop response of the wind farm. The setpoints $P^{u,SPC}_w$ remain below the uncertain wind power $P^{\max}_{w,un}$, demonstrating the uncertainty-awareness of the predictive controller. The error of the predicted wind farm output $P^{y,SPC}_w$  with respect to the open-loop power output $P^{y,ol}_w$, normalized by the mean of $P^{y,SPC}_w$, is 6.5\%. $P^{y,ol}_w$ follows $P^{u,SPC}_w$ reasonably well given that the inner control-loop for the wind farm is well-tuned. The deviation of $P^{y,SPC}_w$ from $P^{u,SPC}_w$ can be attributed to noise in the data collected to calculate $S^*$ in \eqref{eq:Sstar}, the relaxation terms in \eqref{eq:FL}, and the input-output dynamics of the HPP captured in $S^{*}$.  

Figure \ref{fig:SPC_OL_un_Solar_Batt} shows the open-loop response of the solar farm, $P^{y,ol}_s$, and battery, $P^{y,ol}_b$, to the setpoints $P^{u,SPC}_s$ and $P^{u,SPC}_b$, respectively. The error of the predicted solar farm output $P^{y,SPC}_s$  with respect to $P^{y,ol}_s$, normalized by the mean of $P^{y,SPC}_s$, is 8.5\%. For the battery, this value is 10\%.

Figure \ref{fig:SPC_OL_un_load} plots the predicted total power output of the HPP, $P^{SPC}_l=P^{y,SPC}_w+P^{y,SPC}_s+P^{y,SPC}_b$, over the prediction horizon of $N$ samples. The error of $P^{SPC}_l$ with respect to $P^{ol}_l$, normalized by the mean of $P_r$ over the prediction horizon, is 6.1\%. The uncertainty in the maximum available wind power, $P^{\max}_{w,un}$, leads to non-smooth variation in $P^{SPC}_l$ and the total available wind and solar energy $P^{SPC}_{w,un + s}$. The total power setpoint to the wind and solar farms in open-loop is $P^{u,ol}_{w+s}=P^{u,SPC}_w+P^{u,SPC}_s$. Its increasing deviation from $P^{\max}_{w,un + s}$ is undesired since it is beneficial to utilize the available and renewable wind and solar energy before discharging the battery to support the tracking of $P_r$. Given the reasonably low normalized errors in the predictions of the data-driven controller for the open-loop outputs of the power plant components, the predictive controller demonstrated potential in providing its solutions as \textit{forecasts} of HPP power generation.
\begin{figure}[tb]
    \centering
\includegraphics[width=\linewidth]{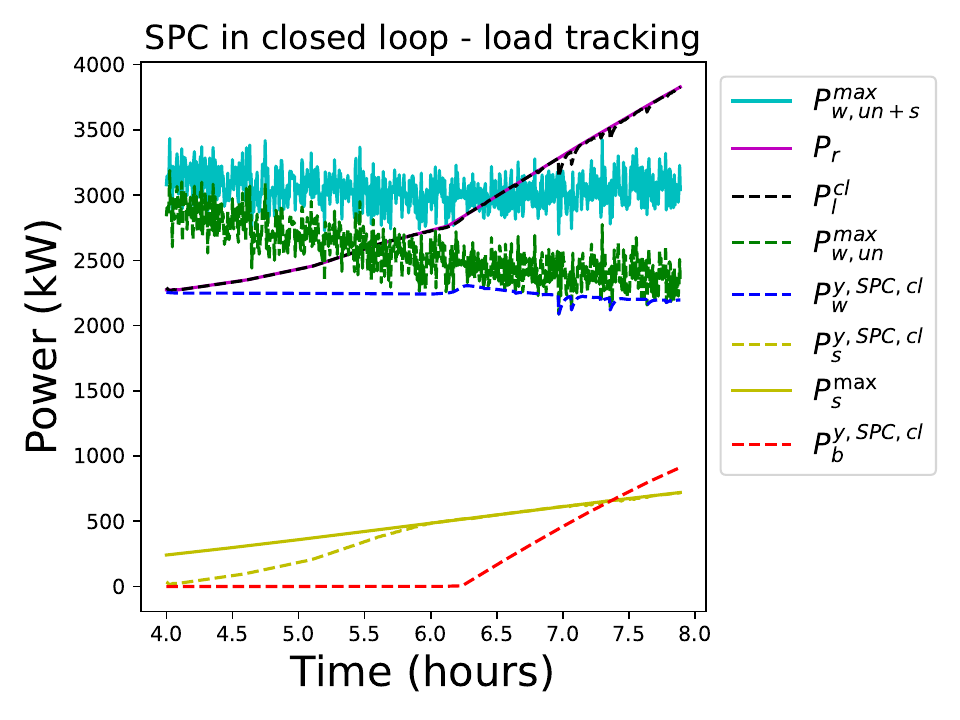}
    \caption{Uncertainty-aware closed-loop data-driven predictive control of the hybrid power plant in Figure \ref{fig:hsc_control_diag}. The plant output $P_l$ tracks the load reference $P_r$ well. The predictive controller anticipates $P_r$  to exceed the maximum available wind and solar energy $P^{\max}_{w,un +s}$ and discharges the battery ($P^{y,SPC,cl}_b>0$) to support load tracking. Maximum available solar energy is utilized ($P^{y,SPC,cl}_s \approx P^{\max}_{s}$) and the wind farm output $P^{y,SPC,cl}_{w}$ is limited by the uncertain available wind power $P^{\max}_{w,un}$.}
    \label{fig:SPC_CL_un_load}
\end{figure}
\begin{figure}[tb]
    \centering
\includegraphics[width=\linewidth]{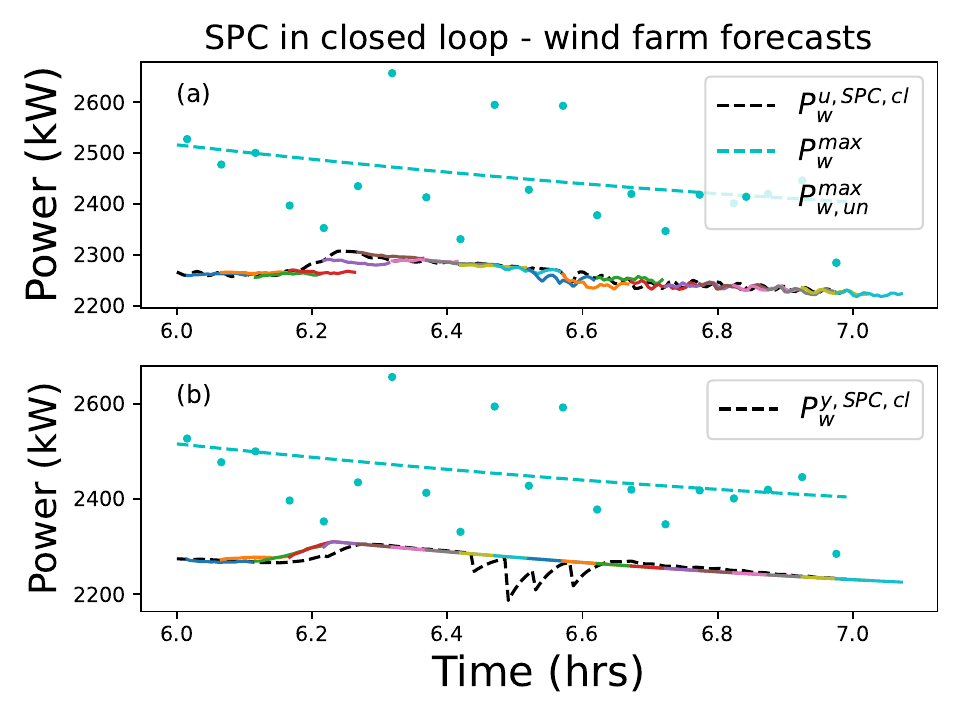}
    \caption{The solution of the uncertainty-aware data-driven predictive controller can be treated as a \textit{forecast} or expected trajectory of the power plant output over a future horizon. The uncertain and available wind power $P^{max}_{w,un}$ is dotted at specific sample times. At these sample times, solid lines represent the prediction of future setpoints, in (a), and wind farm output, in (b). The underlying dashed lines in black represent the setpoints and true output obtained in simulated closed-loop control of the hybrid power plant.}
    \label{fig:SPC_CL_fore}
\end{figure}
\begin{figure}[tb]
    \centering
\includegraphics[width=\linewidth]{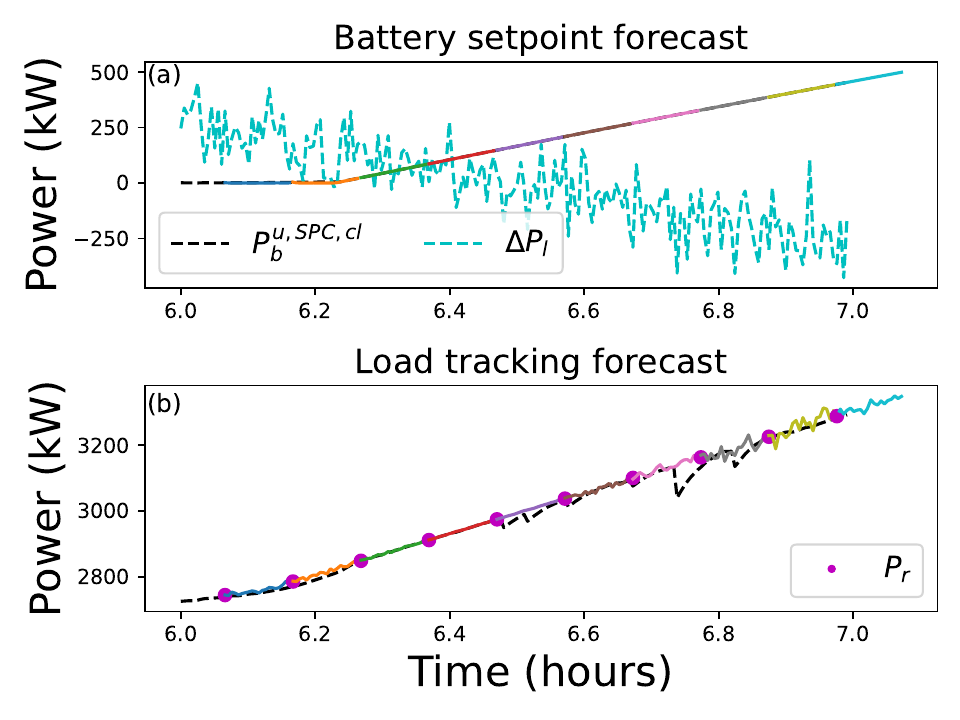}
    \caption{(a) Each solid line is a forecast of the setpoint to the battery, over the prediction horizon of the predictive controller. $\Delta P_l>0$ is the excess ($\Delta P_l>0$) or deficit ($\Delta P_l<0$) in the combined power production of the wind and solar farms, relative to the load reference. In the event of a deficit, the battery must discharge ($P^{y}_b>0$) to support the load. At $\approx$ 6.2 hours, the battery setpoints in solid orange \textit{forecast} the need to discharge and support the load, and turn positive toward the tail of the forecast. This demonstrates the capability of the predictive controller to anticipate and control the state of the battery, despite the presence of uncertainty in the available wind power. (b) Each solid line represents a forecast of the total output of the hybrid power plant given by the predictive controller. This output is obtained by summing the predicted outputs, $P^{y,SPC}_{w}$, $P^{y,SPC}_{s}$, and $P^{y,SPC}_{b}$. Since the true output of the power plant (in dashed black) closely follows the forecasts, the predictive controller is shown to be uncertainty-aware and capable of providing useful forecasts of power generation.}
    \label{fig:SPC_CL_bat_plant_fore}
\end{figure}

\noindent \textbf{Closed-loop control:} Next, the predictive controller is tested in a closed-loop fashion. In closed-loop control, the setpoints selected by the predictive controller are applied to the HPP in a receding horizon fashion. The first setpoint of the solution $u_N$, $u_N^1=\{P^{u,1}_w, P^{u,1}_s, P^{u,1}_b\}$ is applied and the measurement of the resulting output $y=\{P^{y}_w, P^{y}_s, P^{y}_b\}$ is fedback to the controller to solve the optimization again. Noise to signal ratio of zero is considered to isolate the response of the controller to uncertainty in the available wind power.

Figure \ref{fig:SPC_CL_un_load} shows the effective closed-loop tracking of the load reference $P_r$ by the uncertainty-aware predictive controller between 4am and 8am for the test data in Figure \ref{fig:Test_load}. The load reference $P_r$ is tracked well by the HPP load output $P^{cl}_l$. The predictive controller anticipates $P_r$  to exceed the maximum available wind and solar energy $P^{\max}_{w,un +s}$ and discharges the battery ($P^{y,SPC,cl}_b>0$) to support load tracking. Maximum available solar energy is utilized ($P^{y,SPC,cl}_s \approx P^{\max}_{s}$) and the wind farm output $P^{y,SPC,cl}_{w}$ is limited by the uncertain available wind power $P^{\max}_{w,un}$. 

\noindent \textbf{Forecasting power plant output:} Figure \ref{fig:SPC_CL_fore} provides the $N=20$ samples long solution of the predictive controller at evenly spaced sampling intervals between 6am and 7am for the test data in Figure \ref{fig:Test_load}. At each sampling interval, the corresponding (uncertain) upper bound on available wind power, $P^{\max}_{w,un}$, is plotted as a cyan dot. Given that the predictive controller showed potential to predict the open-loop performance of the HPP, its solutions can be explored as \textit{forecasts} for the HPP outputs over a future horizon. 

Figure \ref{fig:SPC_CL_fore} (a) provides forecasts of the setpoints to the wind farm, $P^{u,SPC,cl}_w$, each plotted as a solid line of different color. Figure \ref{fig:SPC_CL_fore} (b) provides forecasts of the power output of the wind farm, $P^{y,SPC,cl}_w$, each plotted as a solid line of different color. All forecasts obey the probabilistic constraint in \eqref{ineq:Pwmxmn} and remain below $P^{\max}_{w,un}$. The underlying black dashed line in the respective plots represents the setpoint and true output recorded in the simulated closed-loop control of the HPP.

Figure \ref{fig:SPC_CL_bat_plant_fore} (a) provides forecasts of the setpoints to the battery, $P^{u,SPC,cl}_b$, each plotted as a solid line of different color. The underlying dashed line in black represents the setpoint utilized in closed-loop control of the HPP. The dashed cyan line represents the excess ($\Delta P_l>0$) or deficit ($\Delta P_l<0$) in power production of the wind and solar farms, relative to the load reference $P_r$. In the event of a deficit, the battery must discharge ($P^{y}_b>0$) to support the load. As seen at $\approx$ 6.2 hours, the battery setpoints in solid orange \textit{forecasts} the need to discharge and support the load, and turns positive toward its tail. This demonstrates the capability of the predictive controller to anticipate and control the state of the battery, despite the presence of uncertainty in the available wind power, in the event of a peak demand from the electricity power grid. Such a capability can provide data, on the future participation of the battery, to the manager of an HPP.

Figure \ref{fig:SPC_CL_bat_plant_fore} (b) provides a forecast of the total output of the HPP with respect to the load reference $P_r$. Each forecast is a solid line of a different color and is obtained by summing the predicted outputs, $P^{y,SPC}_{w}$, $P^{y,SPC}_{s}$, and $P^{y,SPC}_{b}$, over the prediction horizon. The underlying dashed line in black is the true total output obtained in the simulated closed-loop control of the HPP. Given that the true output of the power plant (in dashed black) closely follows the forecasts, the predictive controller is shown to be uncertainty-aware and capable of providing useful forecasts of power generation. This capability can give a manager of an HPP data on the anticipated behavior and output of the plant and assist in HPP management.  
%Point to Table \ref{tab:relaxation_norm}.
\begin{table}[tb]
    \caption{Impact of uncertainty, weights, and quantiles, on the relaxations $\bf{\sigma_y}$ and $\bf{\sigma_u}$ of the behavior constraint in \eqref{eq:FL}. $||\cdot||_F$ is the Frobenius norm.$|\Delta P^{y}_w|$ reflects the change in solution of the predictive controller, for wind power output $P^{y,SPC}_w$, from a baseline scenario of no uncertainty and $\lambda_u$=$\lambda_y$=10. The prediction horizon considered is $N$=20 samples long and begins at 11.88 hours for the load profile $P_r$ shown in Figure \ref{fig:Test_load}.}
    \centering
    \begin{tabular}{l l l l}
        \toprule
        Test & $||\bf{\sigma_u}||_F$ & $||\bf{\sigma_y}||_F$ & $|\Delta P^{y}_w|$\\ 
        \hline
        No uncertainty & 8 kW & 5 kW & 0 kW\\
        Uncertainty, ${\lambda_u}$=${\lambda_y}$= 10 & 38 kW & 17 kW & 115 kW\\
        Uncertainty, ${\lambda_u}$=${\lambda_y}$= 10$^5$ & 1 kW & 0.8 kW & 118 kW \\
        Uncertainty, $q_w$ = -0.4 & 38 kW & 17 kW & 115 kW \\
        Uncertainty, $q_w$ = -1.6 & 152 kW & 66 kW & 660 kW\\
         \bottomrule
    \end{tabular}
    \label{tab:relaxation_norm}
\end{table}
\section{Discussions}
\label{sec:disc}
The previous section showed that the data-driven predictive controller has potential for uncertainty-aware closed loop control of a hybrid power plant. Additionally, in the form of a supervisor, i.e., the $HSC$ in Figure \ref{fig:hsc_control_diag}, the controller is shown to be capable of intelligent decision-making for a) optimal setpoint selection for each HPP component, and b) anticipation and control of the state of battery charge/discharge, despite the presence of uncertainty. Next, we discuss a few research inquiries to retain this predictive controller intelligence, and making the decisions and forecasts trustworthy and explainable.
\begin{enumerate}
    
    \item \textit{How good is my data?}

    The uncertainty-aware predictive controller presented in this work is completely data-driven. Its ability to constrain permissible future trajectories of the system in \eqref{eq:FL} relies on the quality of the HPP input-output measurement data used to define the matrix $S^{*}$ in \eqref{eq:Sstar}. Therefore, it is necessary to develop a quantitative and/or qualitative metric of how well the data $U$ and $Y$ represent the dynamics of the HPP. If this is not the case, the most recent input-output measurements, $u_{T_{ini}}$ and $y_{T_{ini}}$, used in \eqref{eq:FL}, will represent system dynamics different from that encoded by $U$, $Y$, and $S^{*}$. This can lead to

    \begin{enumerate}
        \item     large values of relaxations $\bf{\sigma_u}$ and $\bf{\sigma_y}$ in \eqref{eq:FL} to maintain a feasible optimization, and

        \item incorrect setpoint selection (i.e, decision making) and forecasts of HPP power output.
    \end{enumerate}   

A data-quality metric for noise-corrupted data and data prone to privacy related obfuscation is introduced in \cite{mieth2024data}, while a min-max data-driven optimization technique is introduced in \cite{huang2023robust} for distributional robustness against uncertain data. However, the predictive controller introduced in this work is expected to run in-house, making all required measurement data available and free of obfuscation. The question asked above seeks to determine a metric that provides analytical insight on the appropriateness of the data for predictive control and inform the need to update the data collected, and $S^{*}$.

    \item \textit{Is my data-driven optimization biased?}

    The potential of the data-driven predictive controller to serve as a supervisor for HPP control can lead to an evolution of the objective function and constraints utilized in \eqref{eq:J}-\eqref{ineq:Pbmxmn}. The design of the objective function, constraints, and the tuning of penalties such as $Q_r$, $\lambda_u$, and $\lambda_y$, affect the admissible system outputs, $y_N$, and setpoints, $u_N$, solved by the predictive controller over a prediction horizon. If the gains are tuned too strongly to optimize the objective function or the objective demands an uncharacteristic system trajectory, the constraint in \eqref{eq:FL} may be relaxed excessively to accommodate the solution ($u_N$, $y_N$). This trajectory could be biased toward the objective and may not be representable by $U$, $Y$, and $S^{*}$, in \eqref{eq:Sstar}. An example of how optimization may be optimistically biased is provided in \cite{hobbs1989optimization}.

    \item \textit{Does constraint relaxation of the admissible system trajectories offer any clues?}

    Excessive relaxation of the behavior constraint in \eqref{eq:FL} can lead to increased values of $\bf{\sigma_u}$ and $\bf{\sigma_u}$ in the solution of the predictive controller. An increasing relaxation is a sign of the solution trajectory ($u_N$, $y_N$) being increasingly different than the system dynamics encoded in $U$, $Y$, and $S^{*}$, in \eqref{eq:Sstar}. This may happen due to unaccounted uncertainty in the behavior data (old input-output measurement data of the HPP), incorrectly tuned probabilistic constraints (weather forecasts of high-variance), and improper tuning of the penalties in the objective function.    

    For instance, Table \ref{tab:relaxation_norm} provides the Frobenius norm of the relaxations, $\bf{\sigma_u}$ and $\bf{\sigma_y}$, under different penalties, quantiles $q_w$, and uncertainty. A baseline scenario of no uncertainty (wind speed and available wind power are completely certain) is considered, with $\bf{\sigma_u}$ and $\bf{\sigma_y}$ penalized with $\lambda_y$=$\lambda_u$=10. For a given set of ($\lambda_y$, $\lambda_u$, $q_w$), $|\Delta P^{y}_w|$ reflects the change in solution of the predictive controller, for wind power output $P^{y,SPC}_w$, with respect to the baseline scenario. The prediction horizon considered is $N$=20 samples long and begins at 11.88 hours for the load profile $P_r$ shown in Figure \ref{fig:Test_load}. 
    
    Overall, the presence of uncertainty in the weather forecasts (i.e., wind speed in this study) leads to an increased relaxation of the order of a 100 kW. This can be attributed to the following. The wind farm output is bounded by the probabilistic constraint in \eqref{ineq:Pwmxmn} to remain below the quantile $q_w$ of the uncertain and maximum available wind power, as shown in Figure \ref{fig:SPC_CL_un_load}. The resulting conservative but uncertainty-aware production of wind power differs in behavior from the uncertainty-free input-output measurement data, $U$ and $Y$, collected using feedback-optimization in Figure \ref{fig:Test_load}. Thus, the presence of a probabilistic constraint in \eqref{ineq:Pwmxmn}, to provide uncertainty-awareness, necessitates an increased relaxation of the constraint in \eqref{eq:FL}.
    Applying high penalties and setting $\lambda_y$=$\lambda_u$=10$^5$, lowers relaxation of the behavior constraint. However, this comes at a cost of increased change in solution for the outputs of the solar farm and battery (not reported here for sake of brevity). Lastly, using a more conservative quantile value of $q_w=-1.6$ to set a lower bound on the available wind power in \eqref{ineq:Pwmxmn}, leads to a large change in the trajectory of the wind power output, leading to excessive relaxation of the constraint in \eqref{eq:FL}.

    Therefore, developing a threshold for admissible relaxations can provide insight into a change in underlying dynamics of the HPP. Additionally, such a threshold can point toward a need for improved tuning of penalties in the objective function, and the quantiles for the probabilistic constraints.

    A starting point to determine such a threshold could be determining the relaxation required in the behavior constraint in \eqref{eq:FL} to represent the most recent HPP trajectory, ($u_{T_{ini}}$, $y_{T_{ini}}$). The sequence of measurements ($u_{T_{ini}}$, $y_{T_{ini}}$) will encode the up-to-date input-output dynamics of the HPP. Therefore, larger than expected relaxations, to describe ($u_{T_{ini}}$, $y_{T_{ini}}$), may point toward the need to examine the the optimization framework \eqref{eq:J}-\eqref{ineq:Pbmxmn} for lack of awareness of uncertainty in the data and probabilistic constraints. 

\end{enumerate}
\section{Conclusion}
\label{sec:conc}
This work proposed, and investigated the potential of, an uncertainty-aware data-driven predictive controller as an intelligent decision-maker and supervisor for a hybrid power plant (HPP). Intelligence of the controller is considered and demonstrated in the context of its ability to a) encode dynamics of an HPP through the mere use of available input-output measurement data, b) provide forecasts of the power output of each component of the HPP, and c) anticipate and control the state of charge/discharge of the energy storage device in the event of peak demands from the electricity power grid. The potential of the predictive controller was demonstrated using a simulation study that utilized real-world electricity demand profiles. The controller provided good closed-loop control, and forecasts of setpoints to and outputs of the HPP. Lastly, a few research inquiries to retain intelligence, and make the decisions and forecasts of the predictive controller trustworthy and explainable, are discussed. The next steps involve testing and analyzing the predictive controller's performance using real-world weather forecasts, higher-fidelity plant models, and joint probabilistic constraints to enhance uncertainty-awareness and evaluate its impact on solutions. 

The broader objective of this work is to explore and investigate the utility of the predictive controller as an assistant to the management of an HPP, not only in producing electricity, but also in its participating in electricity markets.

\section*{Acknowledgments}
The work was authored by the Pacific Northwest National Laboratory (PNNL) is operated for the DOE by Battelle Memorial Institute under contract DE-AC06-76RL01830. Funding is provided by the U.S. Department of Energy Wind Energy Technologies Office for project: Path to Nationwide Deployment of Fully Coupled Wind-Based Hybrid Energy Systems. The views expressed in the article do not necessarily represent the views of the DOE or the U.S. Government. The U.S. Government retains and the publisher, by accepting the article for publication, acknowledges that the U.S. Government retains a nonexclusive, paid-up, irrevocable, worldwide license to publish or reproduce the published form of this work, or allow others to do so, for U.S. Government purposes.

\bibliography{aaai25}

% \pagebreak

\end{document}